\documentclass{moriond}

\def\be{\begin{equation}}
\def\ee{\end{equation}}
\def\bea{\begin{eqnarray}}
\def\eea{\end{eqnarray}}

\newcommand{\Photo}{\includegraphics[height=35mm]{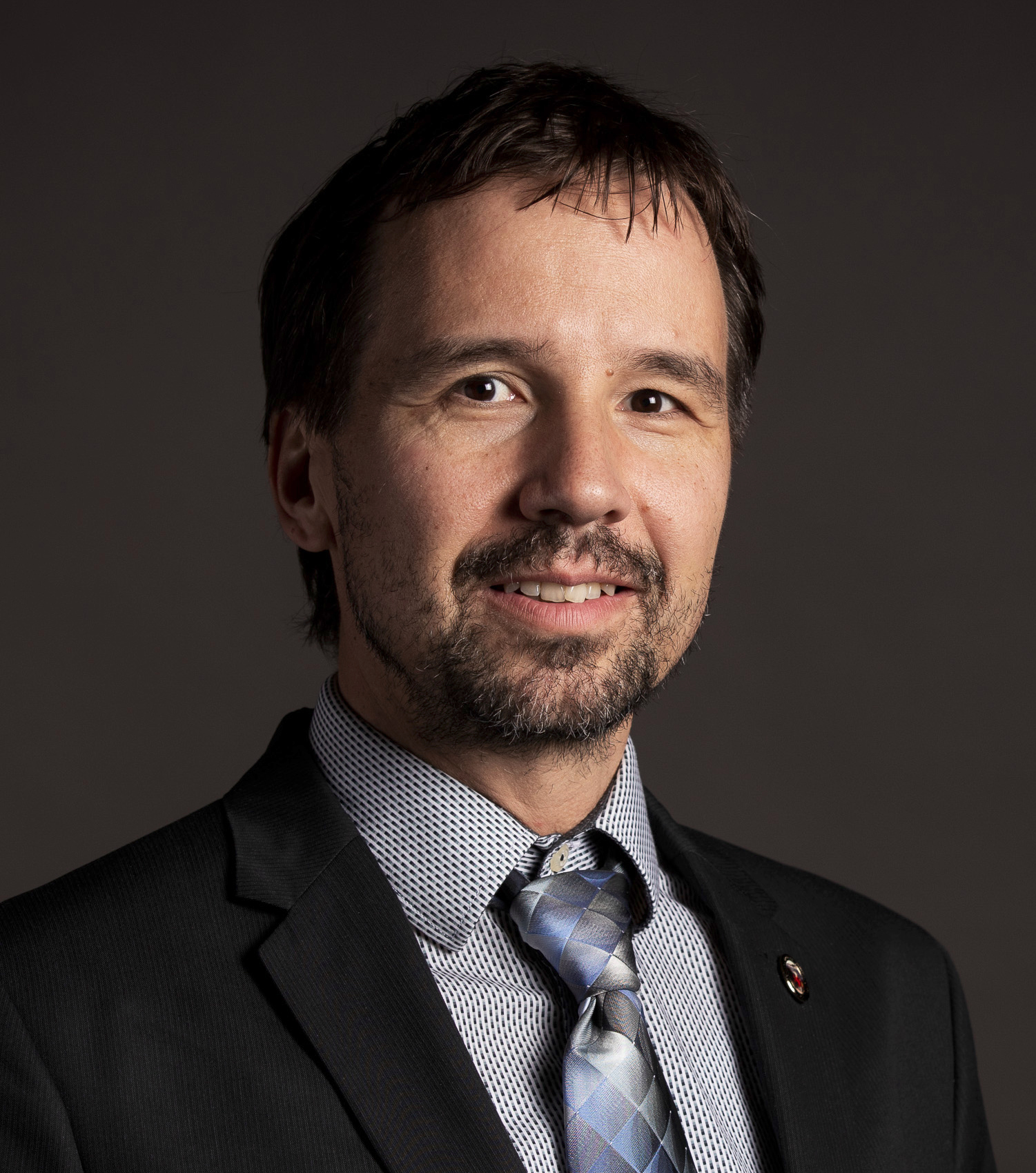}}

\begin{document}
\vspace*{4cm}
\title{Heavy ion physics at CMS and ATLAS: hard probes}

\author{ G\'abor I. Veres on behalf of the CMS and ATLAS Collaborations
\footnotetext{Copyright 2019 CERN for the benefit of the ATLAS and CMS 
Collaborations. CC-BY-4.0 license.}
}

\address{E\"otv\"os Lor\'and University, 
Department of Atomic Physics, \\
P\'azm\'any P\'eter s\'et\'any, 1111 Budapest, Hungary}

\maketitle\abstracts{
Hard probes are indispensable tools to study the hot and dense 
quark-gluon matter created in ultra-relativistic heavy ion collisions. 
These probes are created in the collision itself with a small cross 
section, and they serve as indicators of various properties of the 
medium, such as temperature, viscosity, energy density, transport 
coefficients. Hard probes measured by the CMS and ATLAS experiments at 
the LHC include highly energetic jets and charged particles, quarkonium 
states, and electroweak gauge bosons. An overview of those recent 
experimental results will be given that represent the path towards 
high-precision measurements, even in the challenging, 
high-multiplicity environment created by colliding heavy ions.}

\section{Transport properties, parton energy loss}

In heavy ion physics it is common to compare A+A and p+p interactions in 
order to isolate physical phenomena unique to large colliding systems. 
For the interpretation of such comparisons, it is necessary to quantify 
the modification of parton distribution functions in nuclei, 
including gluon saturation at low $x$. Nuclei can be 
probed with p+Pb collisions at the LHC, and recent results in this area 
include high-energy photons and dijets with a large rapidity separation.

The ATLAS collaboration~\cite{atlaspaper} has recently measured the 
nuclear modification factors, $R_{\rm pA}$, of isolated photons in p+Pb 
collisions, and concluded that the data disfavor a large amount of 
(initial state) energy loss, and impose constraints on the nuclear 
PDFs~\cite{atlasphotons}. Dijets measured in p+Pb collisions with a 
large rapidity separation can probe partons at low x (between $10^{-4} - 
10^{-5}$). No broadening was observed in the azimuthal angle 
correlations for such dijets. Jet pairs, where both jets had a high 
rapidity in the proton-going direction (i.e. sampling low-x partons in 
the Pb nucleus) were found to be suppressed with respect to p+p 
collisions~\cite{atlasdijets}.

\begin{figure}
\centerline{
\includegraphics[width=0.37\linewidth]{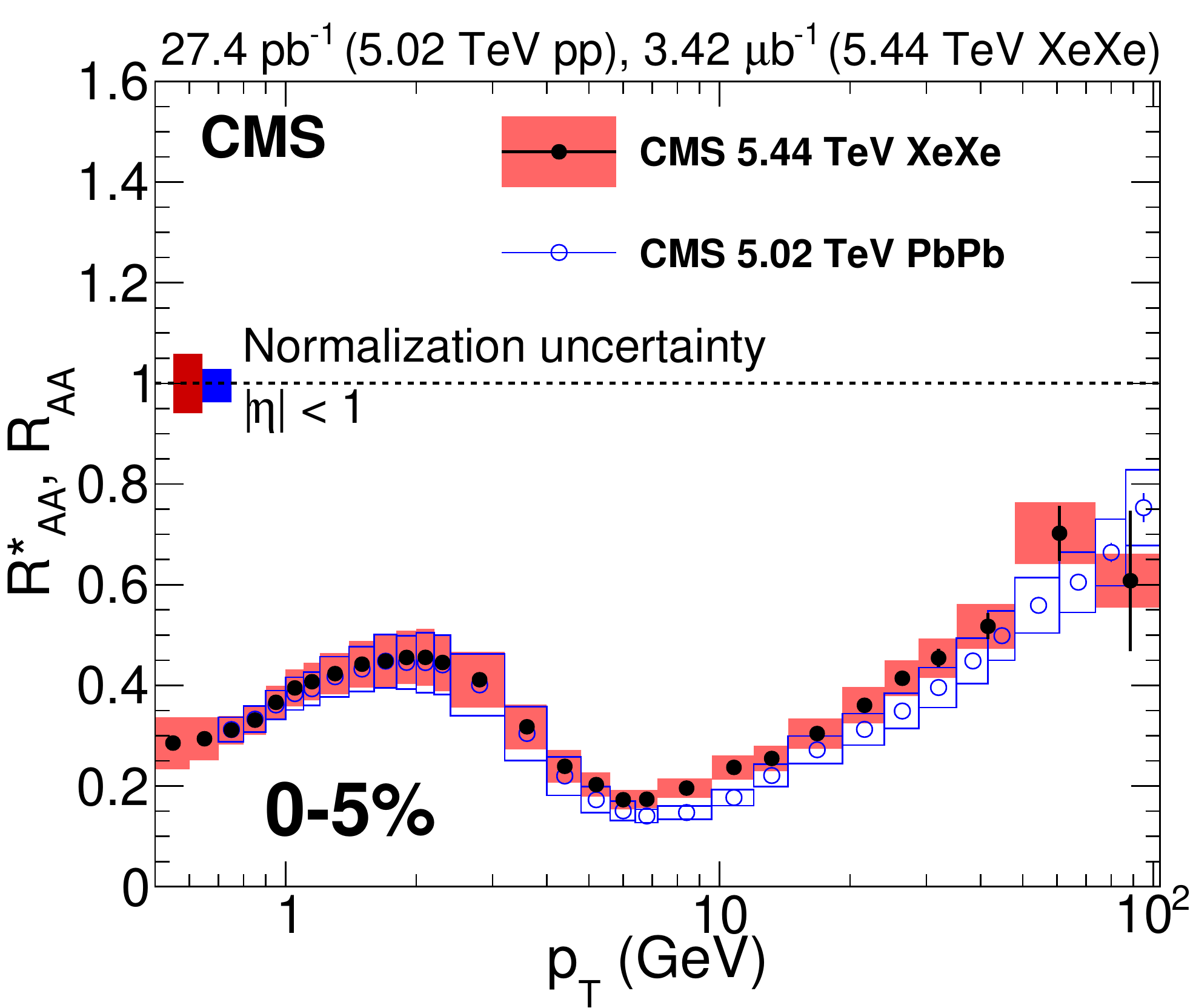}
\includegraphics[width=0.62\linewidth]{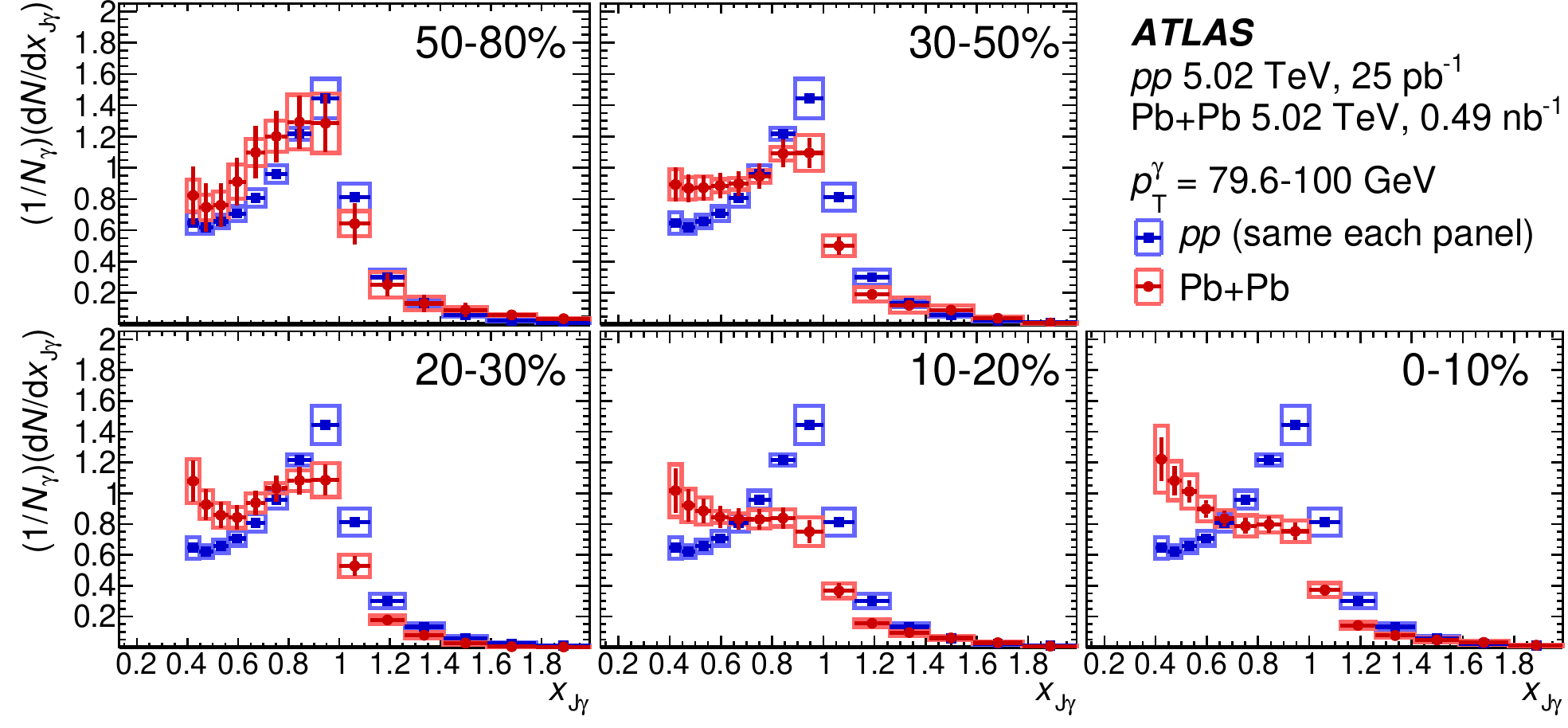}}
\caption{
Left: The charged-particle $R_{\rm AA}$ for Xe+Xe collisions at 
$\sqrt{s_{\rm NN}}=5.44$~TeV for the 5\% most central collisions$^5$,
together with an earlier measurement of $R_{\rm AA}$ in Pb+Pb collisions 
at 5.02~TeV.
Right: photon-jet $p_T$-balance distributions$^7$
in Pb+Pb events (red circles) in 
different centrality bins compared to that in p+p events (blue squares)
for $p_T^\gamma$ = 79.6-100 GeV, where 
$x_J\gamma = p_T^{\rm jet}/ p_T^\gamma$.
}
\label{gammajet}
\end{figure}

Departing from the baseline of nPDFs, one can study final state 
suppression using the nuclear modification factors in Pb+Pb and Xe+Xe 
collisions, as measured by the CMS collaboration~\cite{cmspaper} for 
charged hadrons~\cite{xexe}. On the left panel of Fig.~\ref{gammajet} 
the charged-particle $R_{\rm AA}$ is shown for Xe+Xe collisions at 
$\sqrt{s_{\rm NN}}=5.44$~TeV for the 5\% most central collisions, 
together with earlier data on $R_{\rm AA}$ in Pb+Pb collisions at 
5.02~TeV. The data may indicate a slight difference in suppression at 
high $p_T$. Comparing $R_{\rm AA}$ values at the same number of 
participating nucleons, there is a hint of a greater suppression in 
Xe+Xe collisions, probably due to a geometrical effect.

The final state suppression of jets was also measured by the ATLAS 
experiment with an unprecedented precision recently, showing that the 
nuclear modification factor increases from low to high $p_T$ and from 
central to peripheral Pb+Pb collisions~\cite{atlasjetraa}.
Since photons are not affected by final state interactions, the energy 
loss of jets can be more precisely characterized by selecting photon-jet 
events. The recent data published by the ATLAS 
collaboration~\cite{gammajetatlas} is corrected for accidental pairings 
and unfolded for energy resolution. The result can be seen on the right 
panel of Fig.~\ref{gammajet}, in terms of the distribution of the 
photon-jet $p_T$-balance distributions in different centrality bins, 
where $x_J\gamma = p_T^{\rm jet}/ p_T^\gamma$. One can conclude that 
while many of the jets lose a significant amount of energy
in the most central Pb+Pb collisions, there still remain some
relatively symmetric photon-jet pairs, producing a peak-like 
structure close to unity in $x_J\gamma$.

%----------------------------------------------------------------

\section{Medium temperature, quarkonium states, heavy flavor}

Quarkonium production is a sensitive gauge of the 
temperature in the colored medium created in heavy ion collisions. These 
heavy mesons have a modest binding energy and a large radius, and the 
Debye-screening in the quark-gluon matter may cause their dissociation.
The weakly bound states (like $\Upsilon(2S)$ and $\Upsilon(3S)$) are expected
to suffer a stronger suppression, in comparison to p+p collisions, than more
tightly bound ones, like $\Upsilon(1S)$. The dissociation temperatures
are predicted to be at $2T_c$, $1.2T_c$ and $T_c$ for these three mesons,
where $T_c$ is the critical temperature. 

\begin{figure}
\centerline{
\includegraphics[width=0.49\linewidth]{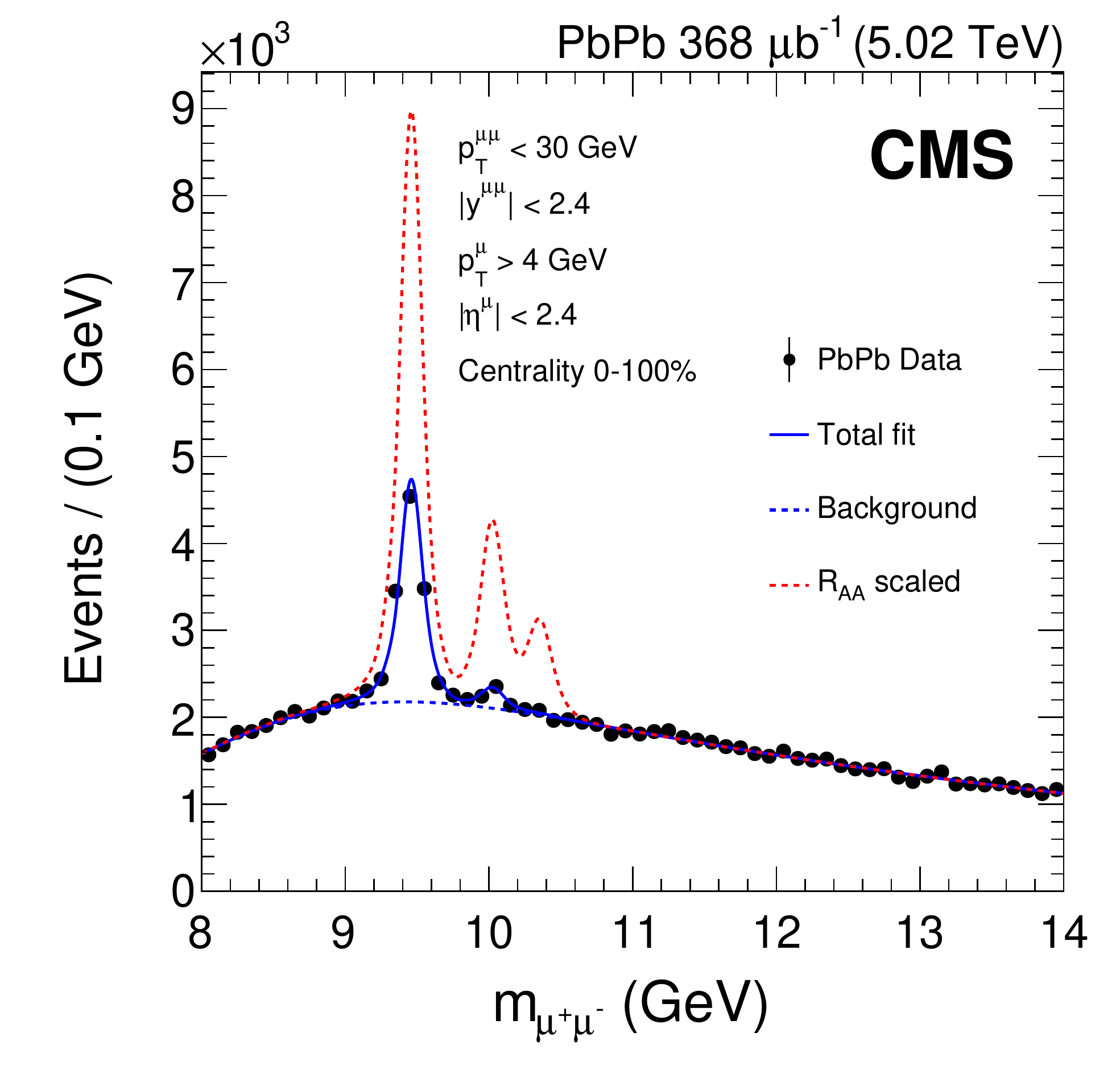}
\includegraphics[width=0.49\linewidth]{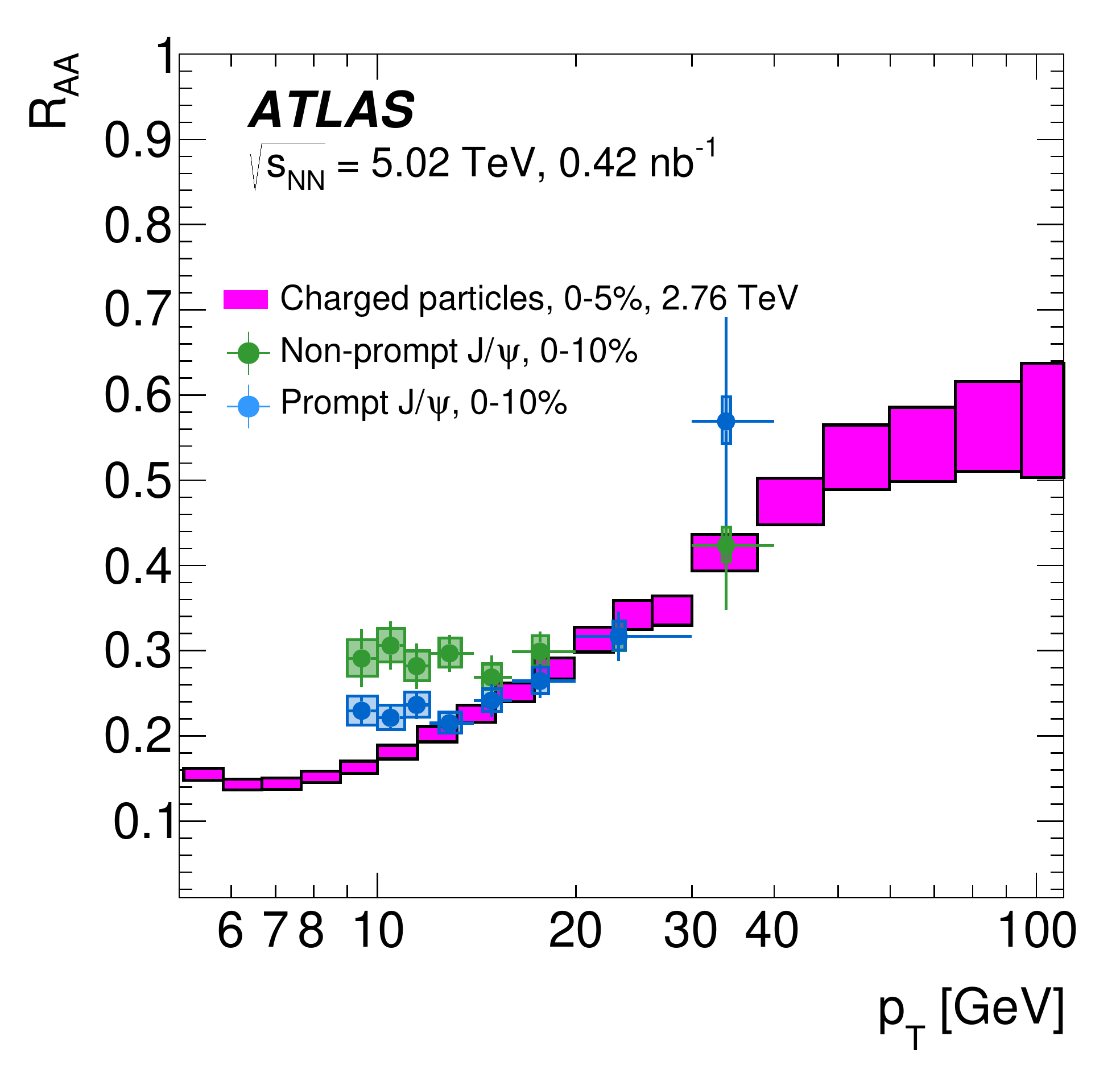}}
\caption{
Quarkonium results in Pb+Pb collisions at $\sqrt{s_{\rm NN}}=5.02$~TeV. 
Left: invariant mass distribution of muon pairs$^8$ for the 
kinematic range $p_T^{\mu^+\mu^-}<30$~GeV and $|y^{\mu^+\mu^-}|<2.4$. 
The result of the fit to the data is shown as a solid blue line. The 
dashed red line is also the result of the same fit but with 
the $\Upsilon$ yield for each state divided by the measured $R_{\rm AA}$.
Right: Comparison of prompt and non-prompt $J/\Psi$ $R_{\rm AA}$ with the 
$R_{\rm AA}$ of charged particles$^{10}$.}
\label{quarkonia}
\end{figure}

Indeed, this successive suppression of the $\Upsilon$ states, measured 
in their dimuon decay channel was observed by the CMS collaboration 
using a high-statistics data set of Pb+Pb collisions~\cite{ycms}. The 
invariant dimuon mass spectrum can be seen on the left panel of 
Fig.~\ref{quarkonia}. The result of the fit to the data including the 
three $\Upsilon$ states and the non-resonant background is shown as a 
solid blue line. The dashed red line represents the result of the same 
fit, but with the $\Upsilon$ yield for each state respectively divided 
by their measured $R_{\rm AA}$ value (i.e. their measured suppression 
with respect to p+p collisions recorded at the same center-of-mass 
energy). The suppression is also found to be gradually strengthening 
with increasing collision centrality.
It was also shown by the recent analysis of the CMS 
collaboration that the excited prompt $\Psi(2S)$ is more suppressed in 
Pb+Pb collisions than the $J/\Psi$ ground state~\cite{cmsjpsi}. 

The $J/\Psi$ mesons also constitute an important tool to characterize 
the b-quark energy loss, since b-decays to $J/\Psi$ can be measured 
separately from prompt $J/\Psi$ production making use of the long 
lifetime of the b quark. The ATLAS collaboration has measured the 
nuclear modification factors of prompt and non-prompt $J/\Psi$ particles 
at high $p_T$, and found a strong suppression in both 
cases~\cite{jpsiatlas}, as can be seen on the right panel of 
Fig.~\ref{quarkonia}. Prompt $J/\Psi$ mesons are suppressed to a similar 
extent as inclusive charged particles, while the non-prompt states 
experience less suppression in the $p_T<20$~GeV range, owing to the more 
modest energy loss of b-quarks compared to light quarks. A further 
confirmation of this phenomena is that the suppression of non-prompt 
$\Psi(2S)$ and non-prompt $J/\Psi$ states, both originating from 
b-decays, were found to be equal. A similar conclusion can be drawn from 
a recent analysis of muons originating from heavy quark decays by the 
ATLAS experiment~\cite{atlasmuons}, and from the CMS measurement of the 
non-prompt $D^0\rightarrow K^-\pi^+$ mesons (coming from b-hadron 
decays), which exhibit significantly less suppression at low $p_T$ 
compared to charged hadrons~\cite{cmsd0}.

It is also interesting to measure the $B_s^0$ state in Pb+Pb collisions 
to test if beauty and strange quarks can coalesce in the environment 
abundant in $s$ quarks, possibly leading to an increase of the 
$B_s^0/B^+$ ratio. The CMS collaboration has published the first result 
on that recently in the $B_s^0\rightarrow \mu^+\mu^- K^+ K^-$ decay 
channel, with a possible indication of such an increase~\cite{bs0}.

%------------------------------------------------------------

\section{Jet substructure}

After considering the spectacular jet quenching (parton energy loss) 
results obtained from heavy ion data, the next immediate question to ask 
is about the possible changes of the jet structure with respect to p+p 
collisions.

\begin{figure}
\centerline{
\includegraphics[width=0.47\linewidth]{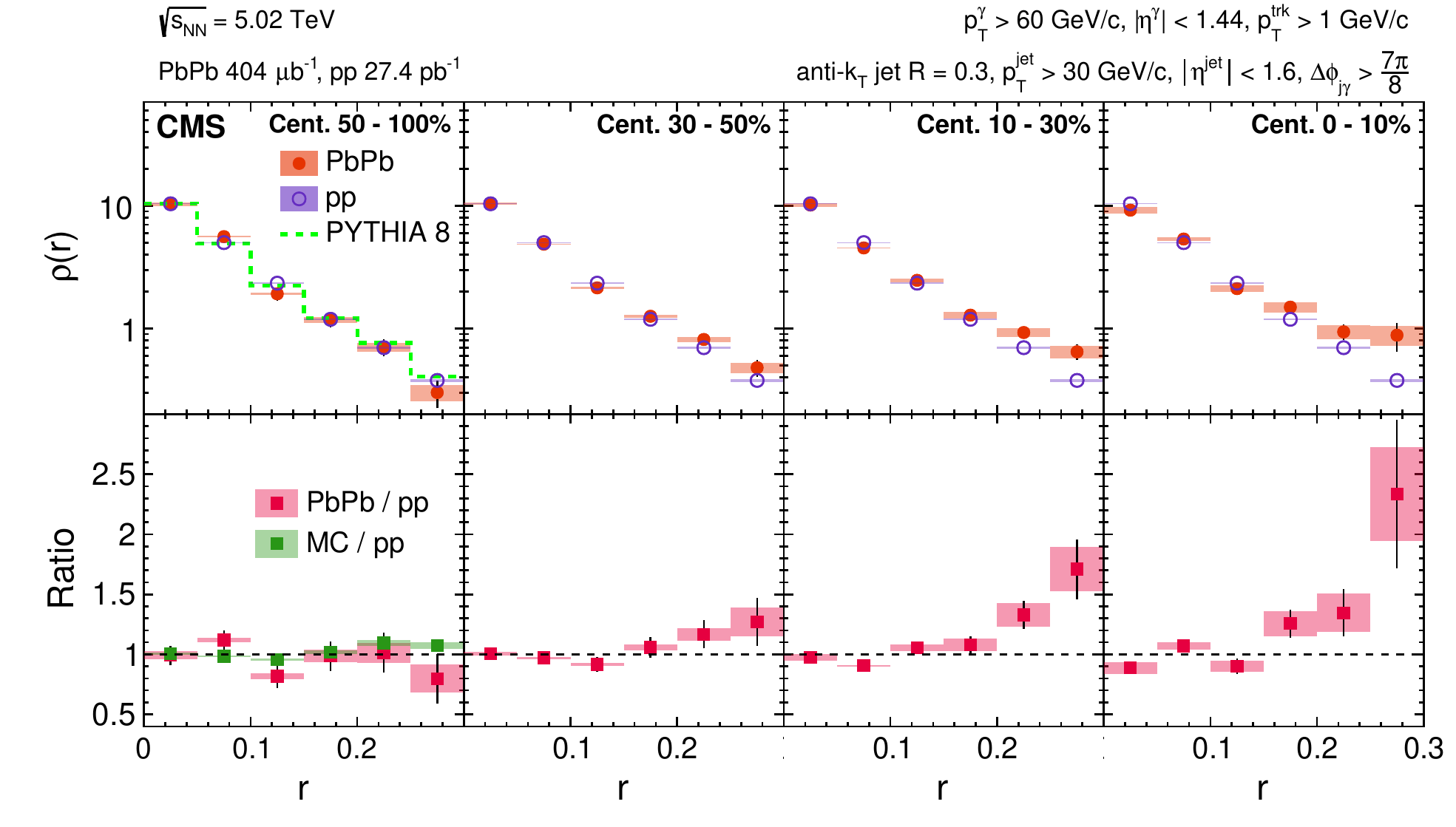}
\includegraphics[trim={0 0 6.245cm 0},clip,width=0.51\linewidth]{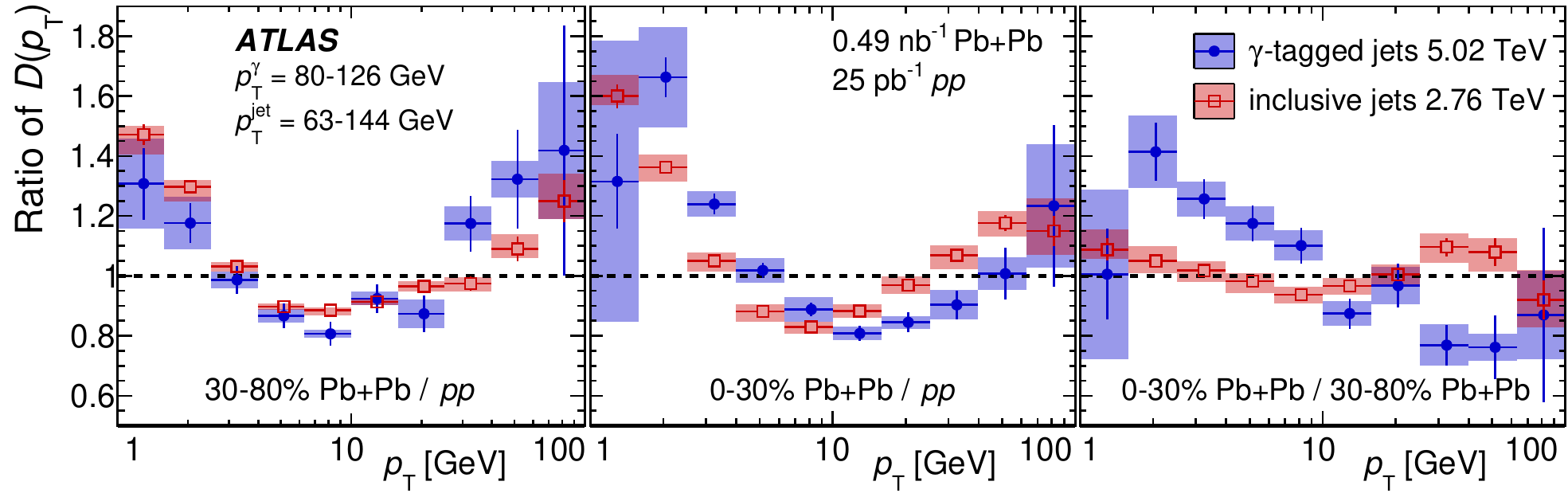}}
\caption{
Left panel, upper plots: the differential jet shape, $\rho(r)$, for jets 
associated with an isolated photon for (from left to right) 50-100\%, 
30-50\%, 10-30\%, 0-10\% Pb+Pb (full circles), and p+p (open circles) 
collisions and from PYTHIA simulation (histogram). Left panel, lower 
plots: the ratios of the Pb+Pb and p+p distributions$^{14}$. Right: ratio of 
the fragmentation function in jets azimuthally balanced by a high-$p_T$ 
photon: 30-80\% Pb+Pb collisions to p+p collisions (left panel) and 
0-30\% Pb+Pb collisions to p+p collisions (right panel). Results are 
shown as a function of charged-particle transverse momentum $p_T$, for 
$\gamma$-tagged jets (this measurement, full markers) and for inclusive 
jets in 2.76 TeV Pb+Pb collisions (open markers)$^{16}$.}
\label{jetshape}
\end{figure}

The CMS experiment has measured the transverse shape (energy density) of 
jets tagged by isolated photons, as a function of the distance $r$ from 
the jet axis for various centrality categories~\cite{jetshapecms}, as 
shown on the left panel of Fig.~\ref{jetshape}. A jet broadening can be 
observed for central Pb+Pb collisions in these photon-jet events, while 
no depletion is visible in the $0.1<r<0.2$ region, as opposed to 
inclusive jets. The longitudinal structure of jets, the fragmentation 
functions, were published by the ATLAS experiment~\cite{atlaslongit}, 
and show that there is an enhancement of particles with a small or very 
large fraction of the jet momentum, and a suppression of particles with 
an intermediate momentum fraction. The jet fragmentation functions were 
also measured in photon-tagged jets~\cite{jetfragatlas}, and a ratio 
with respect to those in p+p collisions are shown on the right panel of 
Fig.~\ref{jetshape}. The jets with a photon partner, dominated by 
quark jets (blue data points), and inclusive jets 
(red points) are modified in a different way in central Pb+Pb events, 
although the interpretation of the data is complicated due to the different 
selection biases.

Finally, the CMS experiment has employed the jet grooming technique to 
remove large-angle, soft radiation, and extract the hard subjets. Since 
the opening angle between those subjets is sensitive to medium induced 
modifications, the distribution of the jet mass of these groomed jets is 
an important observable. These results indicate that available model 
calculations overestimate the yield of jets with a large groomed mass 
relative to the jet $p_T$\cite{jetgroom}.

%\section{Summary}

In summary, the heavy ion research program at the LHC provides a precise 
and detailed set of experimental data, challenging many of the 
phenomenological models concerning jet substructure modifications, and 
supporting various expectations about nuclear PDFs, energy loss of 
light and heavy quarks and quarkonium dissociation.

%\section*{Acknowledgments}

{\bf Acknowledgments.}
The author is grateful for the support of the MTA Momentum program LP 
2015-7/2015 and the support by the grants NKFIA K\_18 128713, NKFIA 
K\_17 124845 and NKFIA FK\_17 123842.

\end{document}